\documentclass[12pt]{iopart}
\usepackage{epsfig}	
\usepackage{bm}

\begin{document}

\def \aj {AJ}
\def \apj {ApJ}
\def \apjl {ApJL}
\def \mnras {MNRAS}
\def \etal {et~al.~}
\def \eg{e.g.}
\def \Section{\S}
\def \spose#1{\hbox  to 0pt{#1\hss}}  
\def \lta{\mathrel{\spose{\lower 3pt\hbox{$\sim$}}\raise  2.0pt\hbox{$<$}}}
\def \gta{\mathrel{\spose{\lower  3pt\hbox{$\sim$}}\raise 2.0pt\hbox{$>$}}}
\def \LCDM {\ifmmode \Lambda{\rm CDM} \else $\Lambda{\rm CDM}$ \fi}
\def \sig8 {\ifmmode \sigma_8 \else $\sigma_8$ \fi} 
\def \OmegaM {\ifmmode \Omega_{\rm M} \else $\Omega_{\rm M}$ \fi} 
\def \OmegaL {\ifmmode \Omega_{\rm \Lambda} \else $\Omega_{\rm \Lambda}$\fi} 
\def \Deltavir {\ifmmode \Delta_{\rm vir} \else $\Delta_{\rm vir}$ \fi}
\def \rs {\ifmmode r_{\rm s} \else $r_{\rm s}$ \fi} 
\def \rrm2 {\ifmmode r_{-2} \else $r_{-2}$ \fi} 
\def \ccm2 {\ifmmode c_{-2} \else$c_{-2}$ \fi} 
\def \cvir {\ifmmode c_{\rm vir} \else $c_{\rm vir}$ \fi} 
\def \cbar {\ifmmode \overline{c} \else $\overline{c}$ \fi}
\def \R200 {\ifmmode R_{200} \else $R_{200}$ \fi} 
\def \Rvir {\ifmmode R_{\rm vir} \else $R_{\rm vir}$ \fi}
\def \v200 {\ifmmode V_{200} \else $V_{200}$ \fi} 
\def \Vvir {\ifmmode V_{\rm  vir} \else  $V_{\rm vir}$  \fi} 
\def  \Vhalo  {\ifmmode V_{\rm halo} \else $V_{\rm halo}$ \fi}
\def \M200 {\ifmmode M_{200} \else $M_{200}$ \fi} 
\def \Mvir {\ifmmode M_{\rm  vir} \else $M_{\rm  vir}$ \fi}  
\def \Mshell  {\ifmmode M_{\rm shell} \else $M_{\rm shell}$ \fi}
\def \Nvir {\ifmmode N_{\rm  vir} \else $N_{\rm  vir}$ \fi}  
\def \Jvir {\ifmmode J_{\rm vir} \else $J_{\rm vir}$ \fi} 
\def \Jshell {\ifmmode J_{\rm shell} \else $J_{\rm shell}$ \fi}
\def \Evir {\ifmmode E_{\rm vir} \else $E_{\rm vir}$ \fi}

\title[High accuracy power spectra]{Dynamical Dark Energy simulations:
high accuracy Power Spectra at high redshift} 

\author{Luciano Casarini$^{1,2}$, Andrea V. Macci\`o$^3$, Silvio
A. Bonometto$^{1,2}$}

\address{$^1$Department of Physics G.~Occhialini -- Milano--Bicocca
 University, Piazza della Scienza 3, 20126 Milano, Italy }
 \address{$^2$I.N.F.N., Sezione di Milano}
 \address{$^3$Max-Planck-Institut f\"ur Astronomie, K\"onigstuhl 17,
 69117 Heidelberg, Germany}


\begin{abstract}
Accurate predictions on non--linear power spectra, at various redshift
$z$, will be a basic tool to interpret cosmological data from next
generation mass probes, so obtaining key information on Dark Energy
nature.  This calls for high precision simulations, covering the whole
functional space of $w(z)$ state equations and taking also into
account the admitted ranges of other cosmological parameters; surely a
difficult task. A procedure was however suggested, able to match the
spectra at $z=0$, up to $k \sim 3\, h$Mpc$^{-1}$, in cosmologies with
an (almost) arbitrary $w(z)$, by making recourse to the results of
$N$--body simulations with $w = {\rm const}$.
In this paper we extend such procedure to high redshift and test our
approach through a series of $N$--body gravitational simulations of
various models, including a model closely fitting WMAP5 and
complementary data. Our approach detects $w={\rm const.}$ models,
whose spectra meet the requirement within 1$\, \%$ at $z=0$ and
perform even better at higher redshift, where they are close to a
permil precision. Available Halofit expressions, extended to
(constant) $w \neq -1$ are unfortunately unsuitable to fit the spectra
of the physical models considered here. Their extension to cover the
desired range should be however feasible, and this will enable us to
match spectra from any DE state equation.

\end{abstract}

\maketitle

\section{Introduction}
\label{sec:intro}

There seems to be little doubt left that a Dark Energy (DE) component
is required, to account for cosmological observables. Its first
evidence came from the Hubble diagram~of SNIa, showing an accelerated
cosmic expansion, but a {\it flat} cosmology with $\Omega_{m} \simeq
0.25$, $\Omega_{b} \simeq 0.04$ and $h \simeq 0.7$ is now required by
CMB and LSS data and this implies that the gap between $\Omega_{m}$
and unity is to be filled by a smooth non--particle component
($\Omega_{m}$, $\Omega_{b}$: matter, baryon present density
parameters; $h$: present Hubble parameter in units of 100 km/s/Mpc;
CMB: cosmic microwave background; LSS: large scale structure; for SNIa
data see \cite{astier,riess}; updated cosmological parameters, taking
into account most available data, are provided within the context of
WMAP5~release \cite{komatsu}.

If DE evidence seems sound, its nature is perhaps the main puzzle of
cosmology. Aside of a cosmological constant $\Lambda$, possibly
related to vacuum energy, and a scalar self--interacting field $\phi$,
various pictures have been recently discussed, ranging from a supposed
back--reaction of inhomogeneity formation to GR modifications and
including even more exotic alternatives (see, {\it
e.g.},~\cite{amendola}).  However, in most of these cases, a DE
component, with a suitable $w(a)$ state parameter ($a$: scale factor),
can still be an {\it effective} description, and a number of
observational projects have been devised, aiming first of all at
constraining $w(a)$ (among them let us quote the DUNE--EUCLIDE
project~\cite{refregier}). Some of them are likely to be realized in
the next decade(s) and, to interpret their outcomes, we need accurate
predictions on selected observables. In particular, it has been
outlined~\cite{huterer} that, to fully exploit weak lensing surveys,
we need predictions on non--linear power spectra, accurate up to $\sim
1 \%$.

$N$--body gravitational simulations safely predict non--linear matter
evolution up to wavenumbers $k \simeq 3\, h\, $Mpc$^{-1}.$ When the
scale of galaxy clusters is approa\-ched, discrepancies from
hydrodynamical simulations, although small, exceed a few
percents~\cite{white,jing,rudd}: too much for the accuracy required.
Within the above range, safe expressions of matter fluctuation spectra
at $z=0$, for any state equation $w(a)$, can be obtained from
simulations of suitable models with $w = {\rm const.}$~. This paper
extends the technique yielding such expressions, so to include higher
redshift $z$, where they will be mostly needed.

A fair approximation to the mass power spectrum for $\Lambda$CDM
models is the {\it Halofit} expression~\cite{smith}, based on the halo
model of structure formation and using numerical simulations to fix
parameters left free by the theoretical analysis. Halofit expressions
were extended ~\cite{mcdonald} to include cosmologies with a constant
state parameter ($ -1.5 < w < -0.5$) and for a fairly wide range of
the parameters $\Omega_c$, $\Omega_b$, $h$, $n_s$, $\sigma_8$ ~(here
$n_s:$ primeval spectral index for scalar fluctuations; $\sigma_8:$
linear r.m.s. fluctuation amplitude on the scale of $8\,
h^{-1}$Mpc). To this aim suitable n--body simulations, in a box of
size $L_{box}=110\, h^{-1}$Mpc and with force resolution
$\epsilon=143\, h^{-1}$kpc, were run.

We tentatively applied such generalized Halofit expressions to our
cases. Unfortunately, soon above rather low $z$'s, they fail to
work. This is not unexpected, as we had been pushing them outside the
expected range of validity. Among the models considered, however,
there are cosmologies closely fitting WMAP5 and complementary
data. Accordingly, further work will be needed to provide suitable
generalized Halofit expressions.

Dynamical Dark Energy (dDE) simulations, with a variable state
parameter $w(a)$ deduced from scalar field potentials admitting a
tracker solutions, have been performed since 2003~\cite{klypinM} (see
also~\cite{maccio,linderJ,solevi}) and compared with $w= const$
simulation outputs. Observables considered in these papers, however,
only marginally included spectra. In a recent work (\cite{francis} FLL
hereafter), however, it has been shown how spectral predictions for
constant--$w$ models can also be used to fit the spectra of
cosmologies with a state parameter
\begin{equation}
\label{poly}
w(a) = w_o + (1-a)w_a~,
\end{equation}
given by a first degree polynomial.

In fact, it had been known since several years that spectra at a given
(low) redshift $\bar z$ mainly depends on the comoving distance
between $\bar z$ and the last scattering band ($d_{LSB}(\bar z)$), at
least for models where baryons and CDM are the only matter components
with a ratio not too far from canonical values~\cite{linderW}.  Aiming
at $1\, \%$ accuracy, FLL seek the constant--$w$ model which grants
the same distance from LSB of an assigned $w(a)$ function, without
varying any other parameter, and claim that spectral differences
between the two models are at the per--mil level for $k < 1\, h$
Mpc$^{-1}$, but still at the percent level for $k < 3\, h$ Mpc$^{-1}$,
at $z=0~.$

On this basis, at $z=0$, 
one could think to use
the {\it Halofit} extension provided
in~\cite{mcdonald} 
to predict the spectra of models with variable $w$, in a large number
of cases.

FLL started from models with a constant equation of state ($w=-K$,
with $K = 0.9, 1, 1.1$), for which the distance from the LSB is
$d_{LSB}^{(K)} (z)$, and then compared them with variable--$w$ models
with the same distance from the LSB at redshift zero:
$d_{LSB}(0)=d_{LSB}^{(K)}(0)$. In the attempt to extend the fit
between spectra from $z=0$ to higher $z$, they just renormalize the
amplitudes of the constant--$w$ model spectra at higher redshift, so
to meet the low--$k$ linear behavior of the variable--$w$ models. This
technique allows them to reduce spectral discrepancies, in average
still below $1\, \%$, but attaining a maximum of a few percents at
larger $k$'s.

Here we shall bypass this renormalization procedure, aiming at the
per--mil precision for any $z$. To this end we extend directly the
criterion suggested in~\cite{linderW} to any redshift, suitably
seeking a constant--$w$ model such that $d_{LSB} (z)/d_{LSB}^{(K)}(z)
\equiv 1$, $z$ by $z$ (allowing, of course, for any value of
$K$). This also requires to follow the variations of model parameters
through their evolution dictated by the assigned $w(z)$. Two options
to do so will be outlined and we shall test the procedure of the
simpler option through a number of $N$--body simulations.

The remainder of this paper is organized as follows. Section
\ref{sec:wvar} is devoted to briefly illustrate such dDE model. In
Section \ref{sec:sims} we present our numerical simulations and their
analysis.  Section \ref{sec:res} contains our results on the Power
Spectrum and a short discussion on them.  Finally in Section
\ref{sec:disc} we present our main conclusions.

\section{The models considered}
\label{sec:wvar}

In this paper we run a series of simulations for two cosmological
models: (i) a model where $w(a)$ is a polynomial (\ref{poly}); (ii) a
SUGRA model. The former model is selected to coincide with one of the
models considered by FLL, so to allow a close comparison of the
outputs.

SUGRA models are an alternative example of faster varying state
equation; they are true dDE model, where DE is a scalar field $\phi$,
self--interacting through a SUGRA potential
\begin{equation}
V(\phi) = (\Lambda^{\alpha+4}/\phi^\alpha)\exp(4\pi \phi^2/m_p^2)
\label{sugra}
\end{equation}
admitting tracker solutions ~\cite{braxM99,braxM01,braxMR}.  Here:
$m_p=G^{-1/2}$ is the Planck mass; $\Lambda$ and $\alpha$ are suitable
parameters; however, in a spatially flat cosmology, once the present
DE density parameter is assigned, either of them is fixed by the other
one. Here we shall define our SUGRA model through the value of
$\Lambda~~ (=0.1$GeV)
\begin{figure}
\begin{center}
\vskip -.3truecm
\includegraphics*[width=12cm]{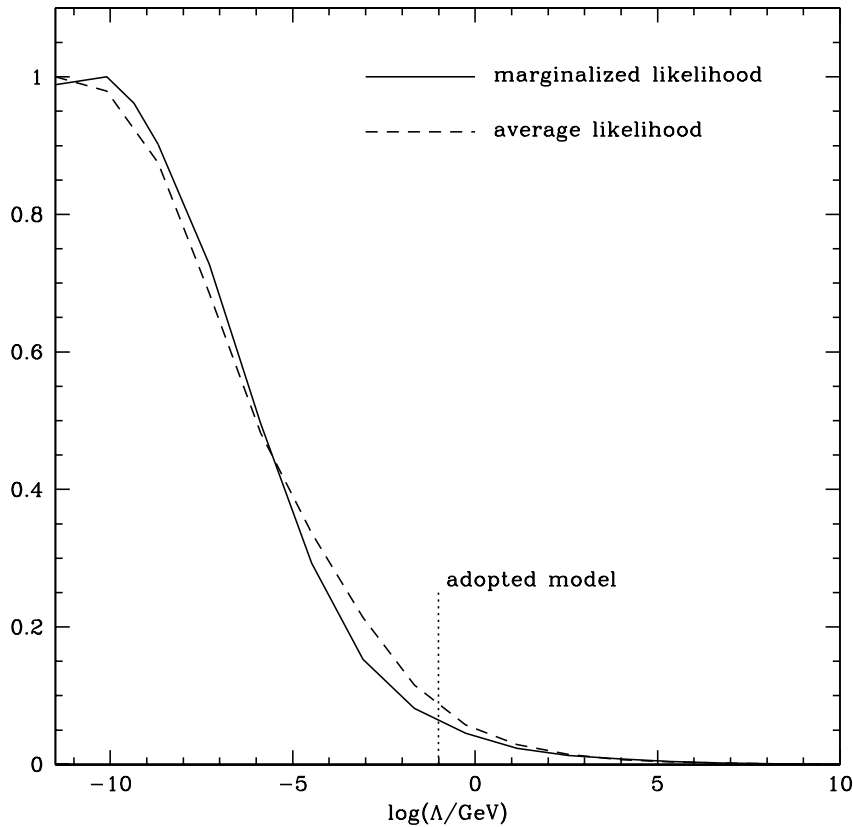}
\end{center}
\vskip -1.truecm
\caption{Likelihood distribution on $\log \Lambda$ for SUGRA
cosmologies.}
\label{wlam}
\end{figure}

\begin{figure}
\begin{center}
\includegraphics*[width=11cm]{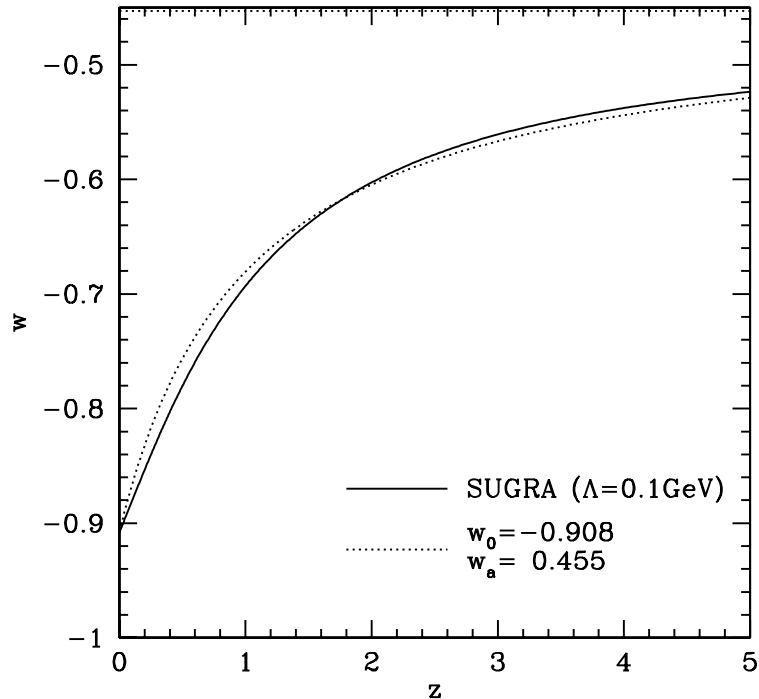}
\end{center}
\vskip -1truecm
\caption{Evolution of the state parameter $w$ for the SUGRA model used
in the simulation ($\Lambda=0.1$), compared with the closest
polynomial $w(a)\, $.}
\label{weps}
\end{figure}

In fact, a fit of SUGRA with data, based on MCMC (MonteCarlo Markov
Chains), was recently obtained by \cite{lavacca}. Data include WMAP5
outputs on anisotropies and polarization \cite{komatsu}, SNIa
\cite{astier} and 2dF data \cite{2df} on matter fluctuation spectra
(including therefore the BAO position). In Figure \ref{wlam} we
exhibit the likelihood distribution on the parameter
$\log(\Lambda/{\rm GeV})$, obtained from the fit.  Using $\Lambda =
0.1~$GeV is a compromise between top likelihood and a physically
significant $\Lambda$~. The values of the other parameters in the
SUGRA simulation are then chosen quite close to the best--fit obtained
once $\Lambda = 0.1~$GeV is fixed, and are reported in Table 1. The
evolution with the redshift of the state parameter for this model is
shown in Figure \ref{weps} and compared with a polynomial $w(a)$ ($w_o
= -0.908$, $w_a=0.455$) coinciding with SUGRA at $z=0$ and $z=10$. In
general, the redshift dependence of the SUGRA state parameter can only
be approached by an expression of the form (\ref{poly}). However, just
looking at the recent WMAP results \cite{komatsu},we see that these
$w_o,~w_a$ are at $\sim 1.5\, \sigma$'s from the best--fit polynomial
parameters.

In Table 1 we give the $z=0$ parameters of the two variable--$w$
models considered.
$$
{\bf Table~~1}
$$$$
\left|~~
\matrix{
    & \Omega_c & \Omega_b & h & \sigma_8 & n_s \cr
{\rm polynom.:} & ~~0.193~~ & ~~0.041~~ & ~~0.74~~ & ~~0.76~~ & ~~0.96 \cr
{\rm SUGRA } & ~~0.209~~ & ~~0.046~~ & ~~0.70~~ & ~~0.75~~ & ~~0.97 \cr
}
~~\right|
$$ 
\vglue .1truecm
Let us remind that the former model is selected mostly for the sake of
comparison. Its polynomial coefficient read $w_o = -0.8$, $w_a =
-0.732$ and it is characterized by a rather low likelihood with respect
to data (see \cite{komatsu})

\begin{figure}
\begin{center}
\includegraphics*[width=9.5cm]{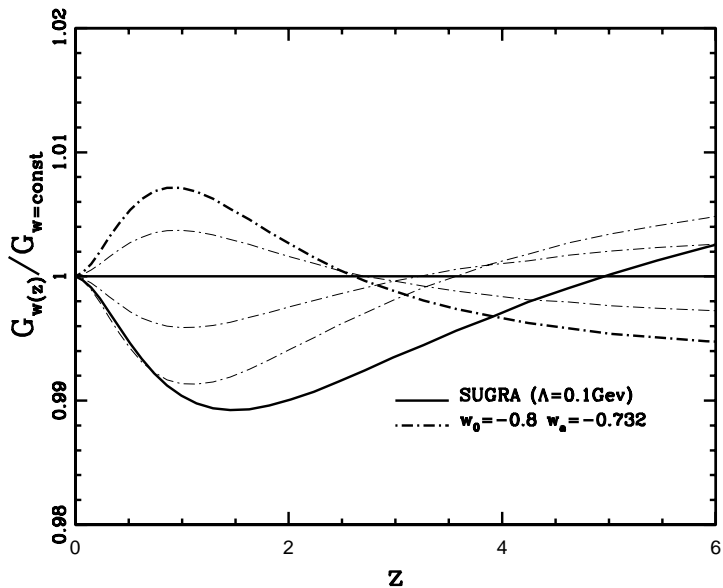}
\end{center}
\vskip -.3truecm
\caption{Linear growth factors for different models, compared with the
growth factor of the constant--$w$ model fitting them at $z=0$.
Besides of the models treated in detail in this work (indicated in the
frame), we also show the growth factors for a further set of models
treated by FLL (see text).}
\label{deltar}
\end{figure}
In Figure \ref{deltar} we show the behavior of the linear growth
factor $G(z)$ for both models, normalized to the growth factor in the
constant--$w$ model yielding the same $d_{LSB}(z=0)$; {\it i.e.}, for
the SUGRA model, $w = -0.7634$; for the polynomial model, as already
mentioned, $w = -1$. For the sake of comparison we report $G(z)$ for a
few other models, whose $d_{LSB}(z=0)$ also coincides with
$\Lambda$CDM (in these models $w_o$ and $w_a$ hold -1.2, -1.1, -0.9
and 0.663, 0.341, -0.359, respectively). $G(z)$ is then the
renormalizing factor for the spectrum at redshift $z$ in the FLL
approach. Spectra worked out with FLL technique will be better where
$G_{w(z)}/G_{w=-1}$ is smaller. We notice that, for each model, there
exist a {\it crossover} redshift $z_{co}$, such that $G(z_{co})$
coincides with the growth factor of $\Lambda$CDM.  At this redshift,
the FLL procedure does not require renormalization.

\section{Variable vs.~constant w}
\label{sec:varia}

\begin{figure}
\begin{center}
\includegraphics*[width=11.5cm]{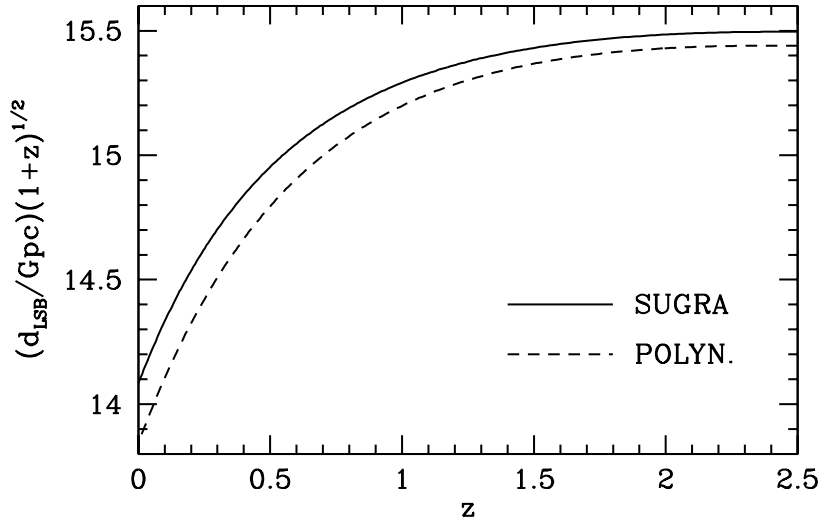}
\end{center}
\vskip -3.truecm
\caption{Distance between $z$ and the LSB in the SUGRA and polynomial models
of Table 1 ($\cal M$ models).}
\label{Dz2}
\end{figure}
\begin{figure}
\begin{center}
\includegraphics*[width=11.5cm]{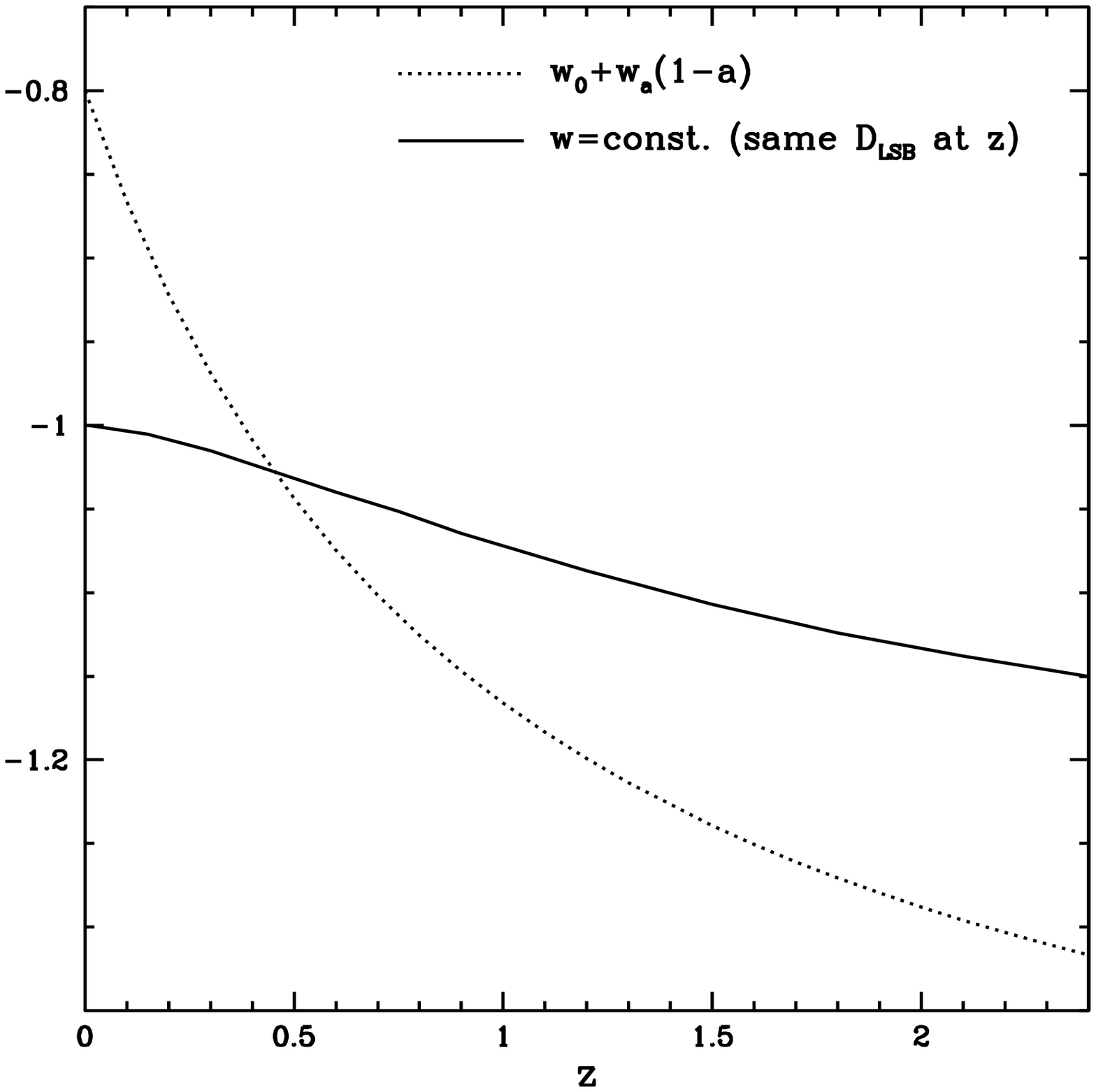}
\vskip -1.4truecm
\includegraphics*[width=11.5cm]{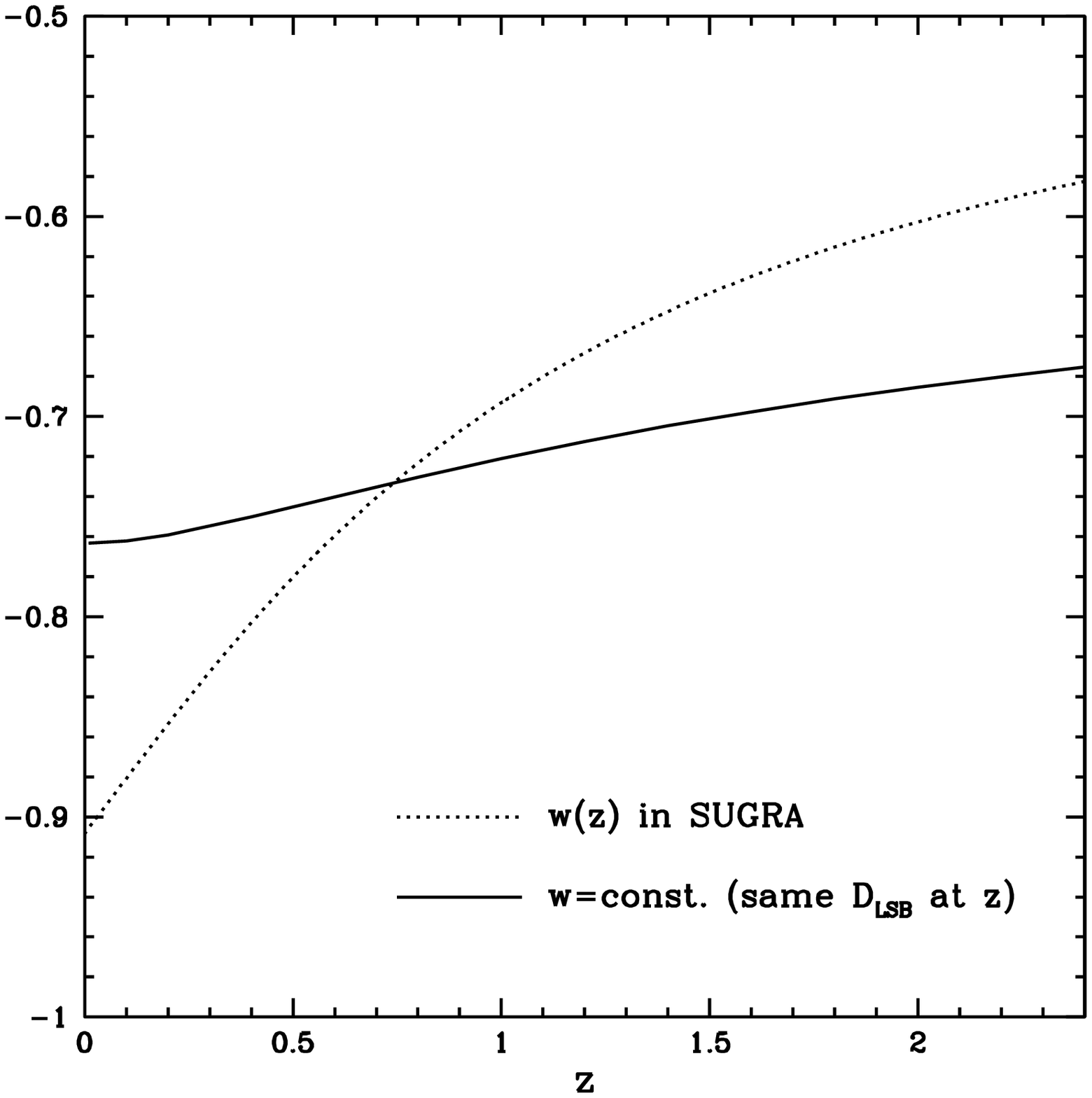}
\end{center}
\vskip -1.truecm
\caption{State parameter of the auxiliary models, as a function of
redshift. For the sake of comparison, also the intrinsic $w(z)$
dependence of each $\cal M$ model is shown.}
\label{wzwz}
\end{figure}
\begin{figure}
\begin{center}
\vskip -2.5truecm
\includegraphics*[width=11.5cm]{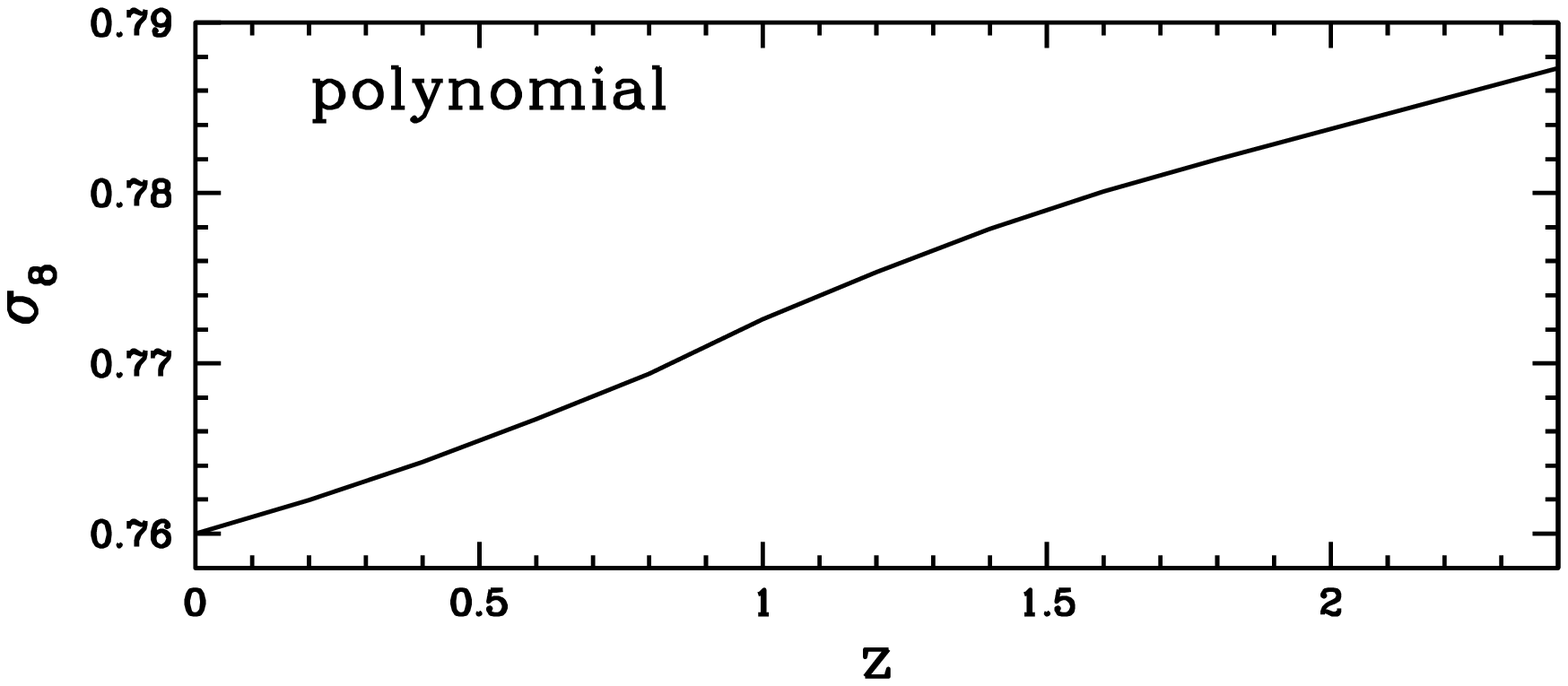}
\vskip -6.4truecm
\includegraphics*[width=11.5cm]{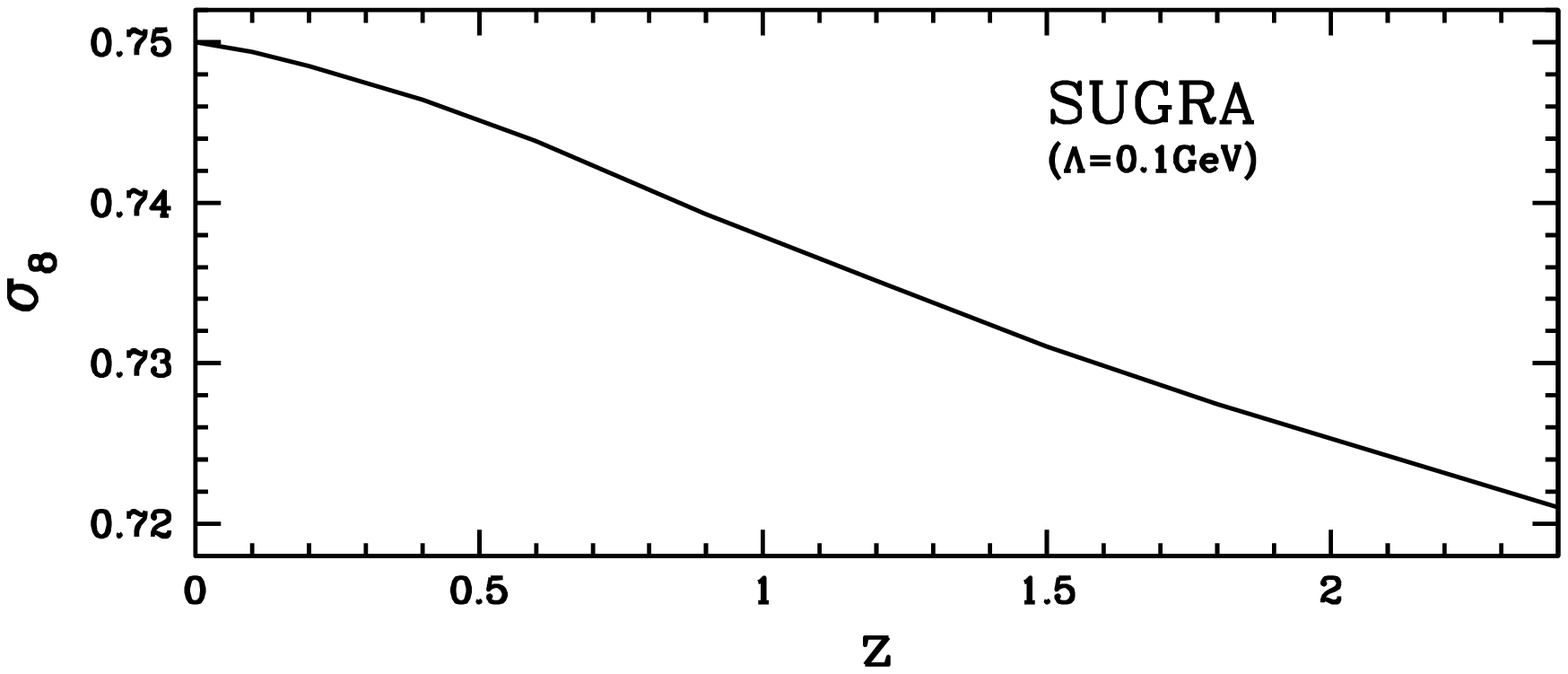}
\end{center}
\vskip -3.truecm
\caption{Values of $\sigma_8$ at $z=0$ for constant--$w$ models whose
r.m.s. fluctuation amplitude meets the $\sigma_8$ value of the
$\cal M$ model at $z$.}
\label{sigz2}
\end{figure}
In order to find a constant--$w$ model whose spectrum approaches dDE
at a redshift $z \neq 0$, one first computes the distance
$d_{LSB}(z)$, from $z$ to the LSB, in the model $\cal M$
considered. At such redshift $z$, cosmological parameters as
$\Omega_b$, $\Omega_c$, $h$, $\sigma_8$ no longer keep their $z=0$
values. However, one can easily calculate them and find a
constant--$w$ model whose parameters at that $z$ are $\Omega_b(z)$,
$\Omega_c(z)$, $h(z)$, $\sigma_8(z)$ and coincide with the values in
$\cal M$. In such model, then, the $z=0$ values of the parameters will
be different from their values in $\cal M$. Such model is then
expected to have a non--linear spectrum closely approaching $\cal M$
at $z$, just as FLL found at $z=0$.  The models defined in this way
shall clearly be different for different $z$ and will be said to
satisfy to the {\it strong requirement}.

Among other difficulties, making use of such models to approximate
high--$z$ spectra causes a technical problem which stands on the way
to test this approach through simulations. Let $L$ be the side of the
box to be used. As already outlined, for the $\cal M$ and
constant--$w$ models, the $z=0$ values of $h$ are different;
therefore, if we take equal values for $Lh$ (or $L$ measured in
$h^{-1}$Mpc) we shall have different $L$ values. A similar problem
occurs when we normalize, as the r.m.s.  fluctuation amplitude
$\sigma_8$ refers to the $h$ dependent scale of $8\, h^{-1}$Mpc.  All
that induces quite a few complications, if we aim at comparing fairly
normalized spectra of different models, with the same seed and the
same wavenumber contributions.

This difficulty can be soon overcame if we replace the above strong
requirement with the {\it weak requirement} we shall now define. Let us
first notice that, being
\begin{equation}
H^2 = {8 \pi \over 3} G \rho_{cr} = {8 \pi \over 3} G \rho_{m}
\Omega_m^{-1}
\label{hz}
\end{equation}
at any redshift, $\Omega_m h^2$ however scales as $a^{-3}$,
independently of DE nature. Accordingly, a constant--$w$ model whose
$z=0$ values of $\Omega_b$, $\Omega_c$ and $ h^2$ coincide with those
of $\cal M$, will share with it the values of the reduced density
parameters $\omega_b = \Omega_b h^2$ and $\omega_c = \Omega_c h^2$ at
any redshift (attention is to be payed, all through this paper, to the
different meanings of the symbols $w$ and $\omega$, respectively state
parameter and reduced density parameters). In order to have the same
$d_{LSB}(z)$, of course, it shall have a different $h(z)$, that we
however do not need to evaluate explicitly. What we need to know are
the value of the {\it constant} $w(z)$ as well as the value $\sigma_8
(z)$, to be assigned to the r.m.s. fluctuation amplitude at $z=0$, to
meet the $\cal M$ value of $\sigma_8$ at $z$.

In this paper we shall test this {weak requirement} (W.R.) that the
dDE model $\cal M$ and the {\it auxiliary} model $\cal W$$ (z)$ have
equal $d_{LSB}(z)$, $\sigma_8$, $\omega_c$ and $\omega_b$, while $h$
is chosen with an {\it ad--hoc} criterion, yielding a $z$--independent
$h$ at $z=0$.

In Figures \ref{Dz2}, \ref{wzwz}, \ref{sigz2} we report the distance
from the LSB $d_{LSB}(z)$, as well as the values of $w(z)$ and
$\sigma_8(z)$ for the auxiliary models $\cal W$$(z)$ defined according
to the W.R.~.

FLL had been seeking an auxiliary model only at $z=0$, making recourse
to a different treatment at greater redshift, claimed to grant a
precision $\cal O$$(2\, $--$3 \%)$. As we shall see, seeking an
auxiliary model at any redshift, according to the W.R., allows a
precision $\sim 10$ times better in the relevant $k$ range and does
not lead to numerical complications. The goal would be complete if
such models were in the parameter ranges considered by generalized
Halofit expressions~\cite{mcdonald} (at $z=0$):
$0.211<\Omega_m<0.351$, $0.041<\Omega_b<0.0514$, $0.644<h< 0.776$,
$0.800<\sigma_8<0.994 $, $0.915<n_s<1.045$.

The selected $\Omega_c$, $\Omega_b$, $h$, $n_s$ are fine. Figures
\ref{wzwz} show that the requirement is met also by $w$.  On the
contrary, Figures \ref{sigz2} show that the $\sigma_8$ value, tuned to
observational data, lays outside the range considered at any $z$, for
both $\cal M$ models.

This is most unfortunate. As we shall see the generalized expression
in \cite{mcdonald} will be scarcely useful to the present aims.

\section{Numerical simulations}
\label{sec:sims}

We shall compare simulations starting from realizations fixed by
identical random seeds. Transfer functions generated using a modified
version of the {\sc CAMB} package~\cite{lewis} are used to create
initial conditions with a modified version of the PM software by
Klypin \& Holzmann~\cite{klypinH}, able to handle different
parameterizations of DE~\cite{klypinM,mainini}. (Possible contributions from DE
clustering on super--horizon scales are ignored in this work).
Simulations were made by using two different programs: the {\sc art}
code~\cite{kravtsov}, courtesy of A. Klypin, and the {\sc pkdgrav}
code~\cite{stadel}, which has been modified to deal with any variable
$w(a)$ for this work. The {\sc art} code has been mainly used as
a benchmark to test our modified version of {\sc pkdgrav} (see
appendix A); in the following we will only present results obtained
with this latter code.

An important issue, when high accuracy is sought, is a suitable
handling of cosmic (sample) variance. In order to address this point
we create a series of simulations sharing the same box size and
particle number, but with different realizations of the initial random
density field. 

The reference point are the simulations of the $\cal M$ models,
performed in a box with side $L_{box}=256 h^{-1}$Mpc, a particle
number $N=256^3$ and a gravitational softening $\epsilon = 25 h^{-1}$
kpc. We create 4 different realizations of them, only differing in the
random seed used to sample the phase space. These initial conditions
are then evolved from $z=24$ to $z=0$ and nine outputs are saved, at
redshift $z_k$, from $z_0 = 2.4$ to $z_8 = 0$. For each redshift we
then have the parameters, at $z=0$, of the corresponding auxiliary
$\cal W$$(z_k)$ model, as shown in Section \ref{sec:varia}. By using
them we create suitable initial conditions for $\cal W$$(z_k)$ models,
using the same random seed of the corresponding $\cal M$ model, and
evolve the $\cal W$$(z_k)$ model down to $z_k$ (9 redshift values
$z_k$, 9 auxiliary models $\cal W$$(z_k)$ for each random seed). For
instance, in order to build a spectrum expected to coincide with SUGRA
at $z=1.2$, we run a simulation of a constant--$w$ model with $w =
-0.7124$ and $\sigma_8 = 0.7351$ at $z=0$.

\subsection{Halos in simulations}
\label{ssec:analy}

\begin{figure}
\begin{center}
\includegraphics*[width=14.5cm]{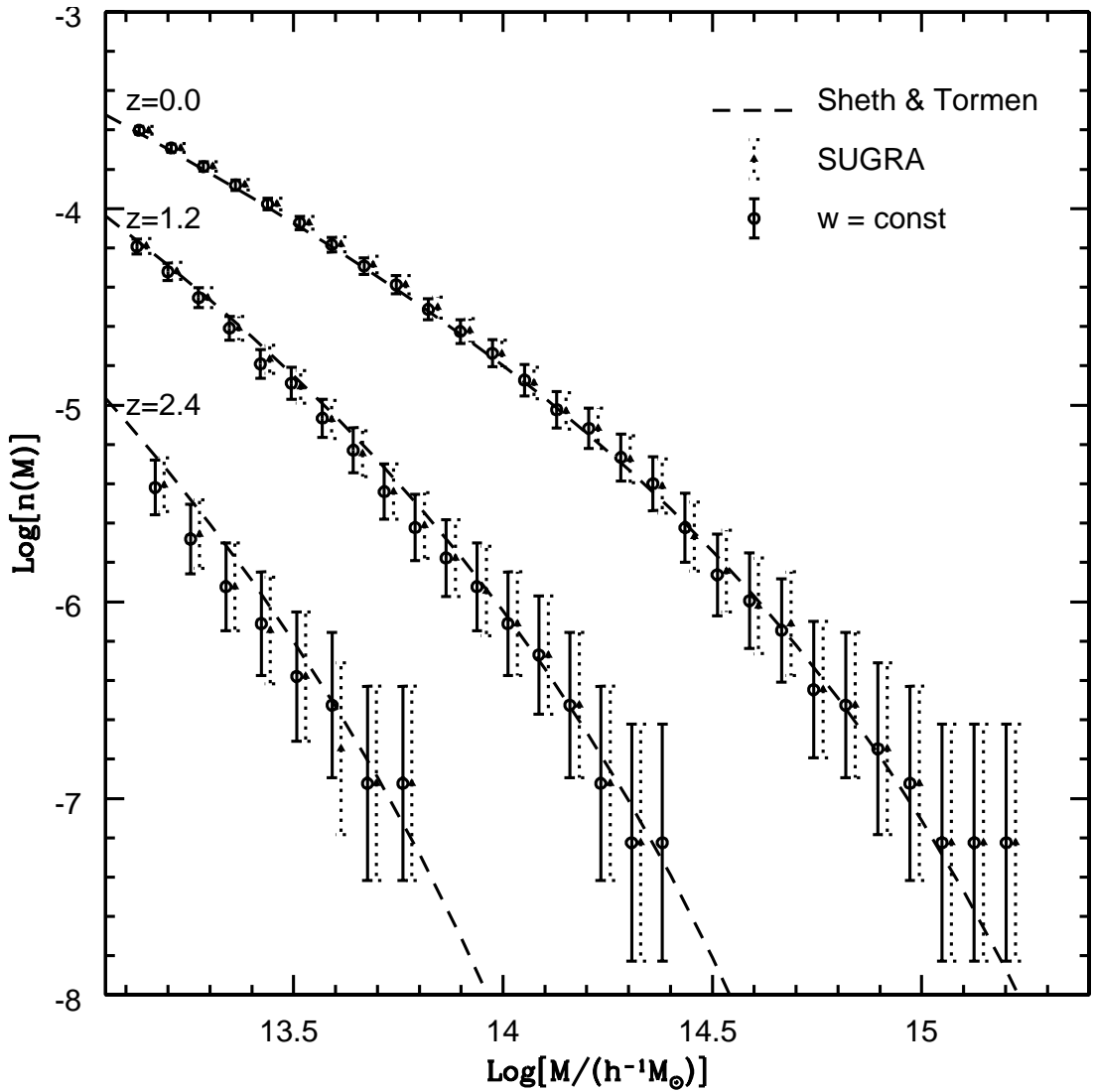}
\vskip -2.truecm
\end{center}
\vskip -.4truecm
\caption{Number of halos per $(h^{-1}$Mpc$)^3$ in the SUGRA model
considered and in the corresponding auxiliary $\cal W$ models.
Results are practically undiscernable and in agreement with Sheth \&
Tormen expressions. The mass functions for the polynomial model behave
quite similarly.}
\label{fmass}
\end{figure}
Before reporting results on spectra and their evolution, we wish to
exhibit a few results on halo formation. Since ever, mass functions
have been considered a basic test for simulations. This is also an
independent test going beyond spectral fits: spectra are related just
to 2--point functions and their fitting formulae derive from
theoretical elaborations on 2--point correlations; mass functions,
instead, as the halo concentration distribution or the void
probability, depend on convolutions of n--point correlations.

We look for virialized halos using a Spherical Overdensity (SO)
algorithm.  Candidate groups with a minimum of $N_f=200$ particles are
selected using a FoF algorithm with linking length $\phi = 0.2 \times
d$ (the average particle separation).  We then: (i) find the point $C$
where the gravitational potential is minimum; (ii) determine the
radius $ r$ of a sphere centered on $C$, where the density contrast is
$\Deltavir$, with respect to the critical density of the Universe.
Using all particles in the corresponding sphere we iterate the above
procedure until we converge onto a stable particle set.  For each
stable particle set we obtain the virial radius, $\Rvir$, the number
of particles within the virial radius, $\Nvir$, and the virial mass,
$\Mvir$.  We used a time varying virial density contrast $\Deltavir$,
whose value has been determined, according to linear theory, by using
the fitting formulae in~\cite{mainini}.  We include in the halo
catalogue all the halos with more than 100 particles.

Mass functions are consistent with Sheth \& Tormen predictions at all
$z$, almost always within 2$\sigma$'s (Poisson errors), while
differences between $\cal M$ and $\cal W$ models can hardly be
plotted. This allows us to formulate the conjecture that also higher
order correlation functions coincide. As future cosmic--shear surveys
will enable to inspect the 3--point function (and possibly to detect
some higher order signal), we plan to compare n--point functions in
$\cal M$ and $\cal W$ models, in a forthcoming paper.

\section{Results on the Power Spectrum}
\label{sec:res}
\begin{figure}
\begin{center}
\includegraphics*[width=11cm]{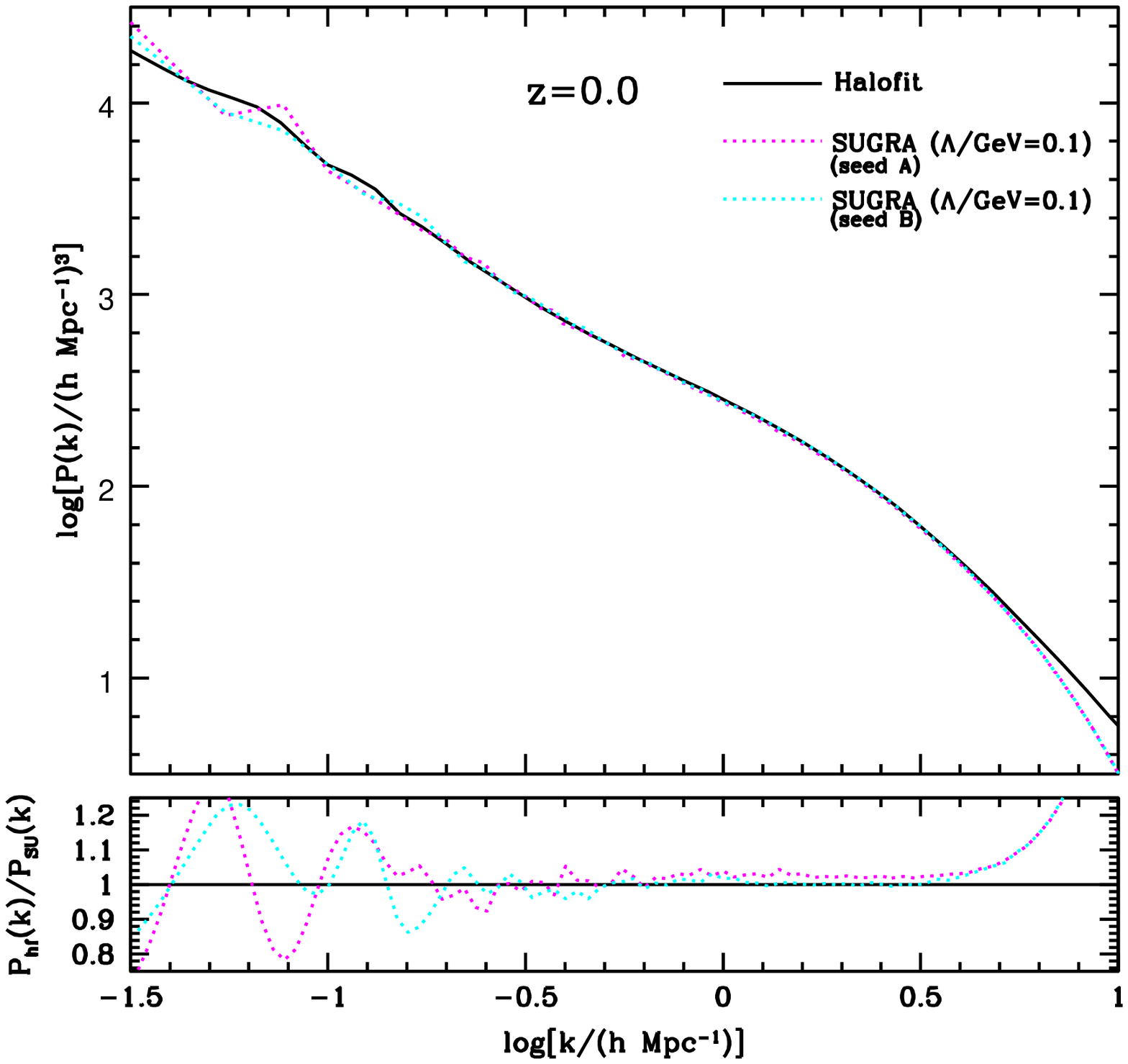}
\vskip -1.truecm
\includegraphics*[width=11cm]{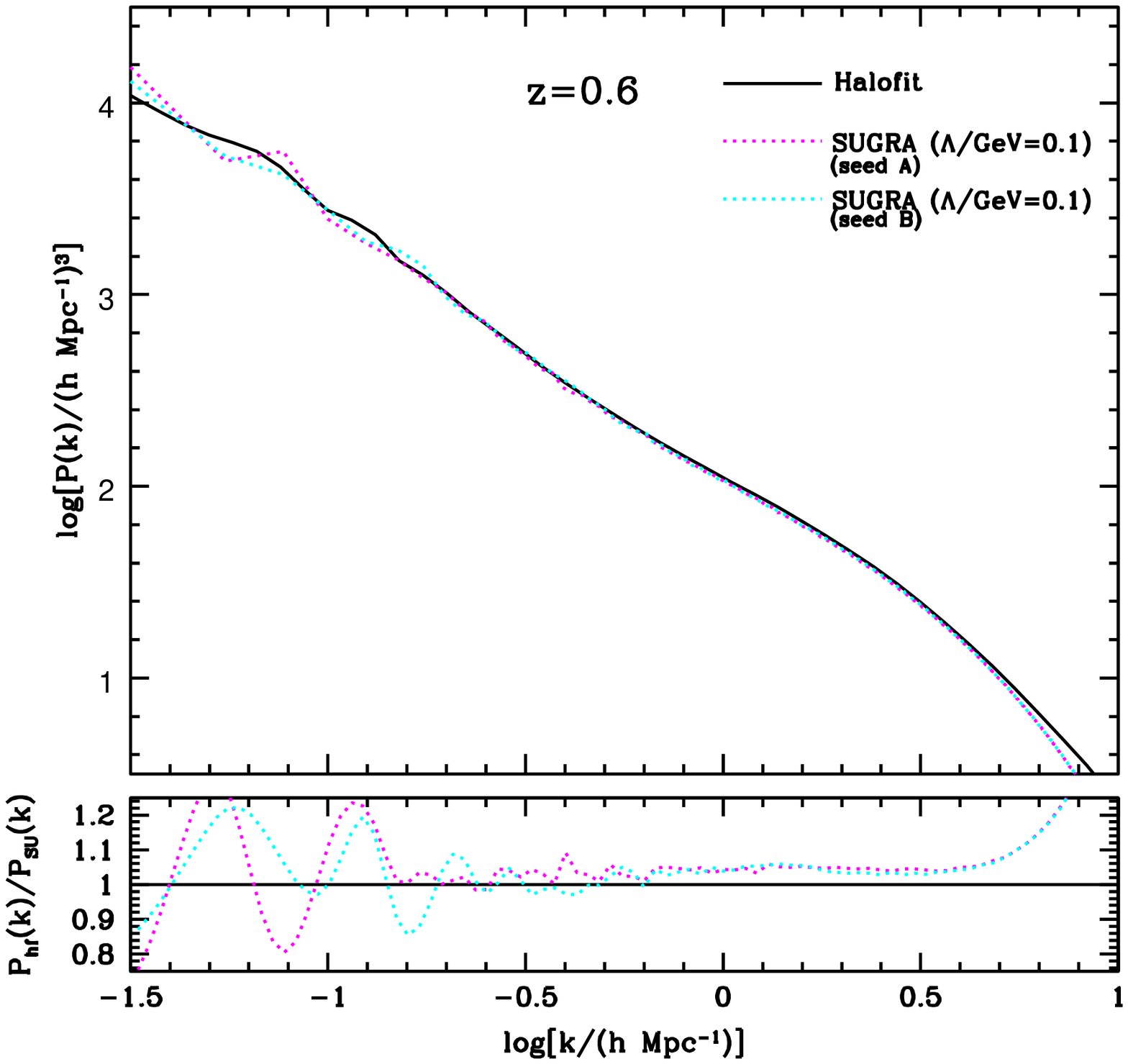}
\end{center}
\vskip -1.truecm
\caption{Spectra of SUGRA and auxiliary models at $z=0$ and 0.6.  Two
realizations are shown. Differences between $\cal M$ and $\cal W$
models are so small to be unappreciable in these plots. In the lower
frames, we rather exhibit the ratio with a Halofit expression, as well
as the difference between realization. At $z=0$, where $\cal M$ model
spectra are quite close to $\Lambda$CDM, this is of the same order of
the discrepancy from Halofit. At $z=0.6$ the discrepancy from a Halofit
expression already overcomes $5\, \%~$.}
\label{s00s04}
\end{figure}
\begin{figure}
\begin{center}
\includegraphics*[width=11cm]{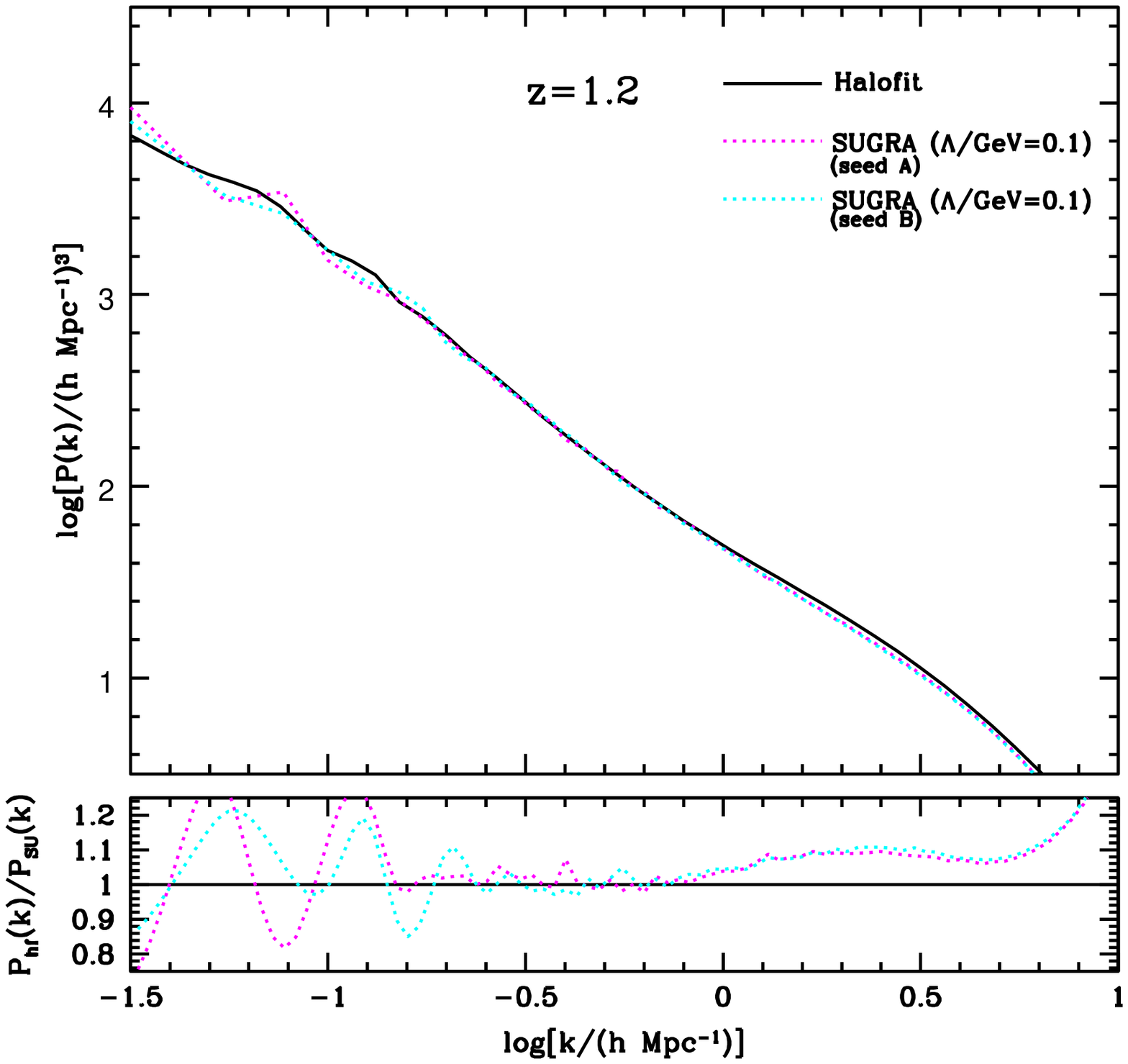}
\includegraphics*[width=11cm]{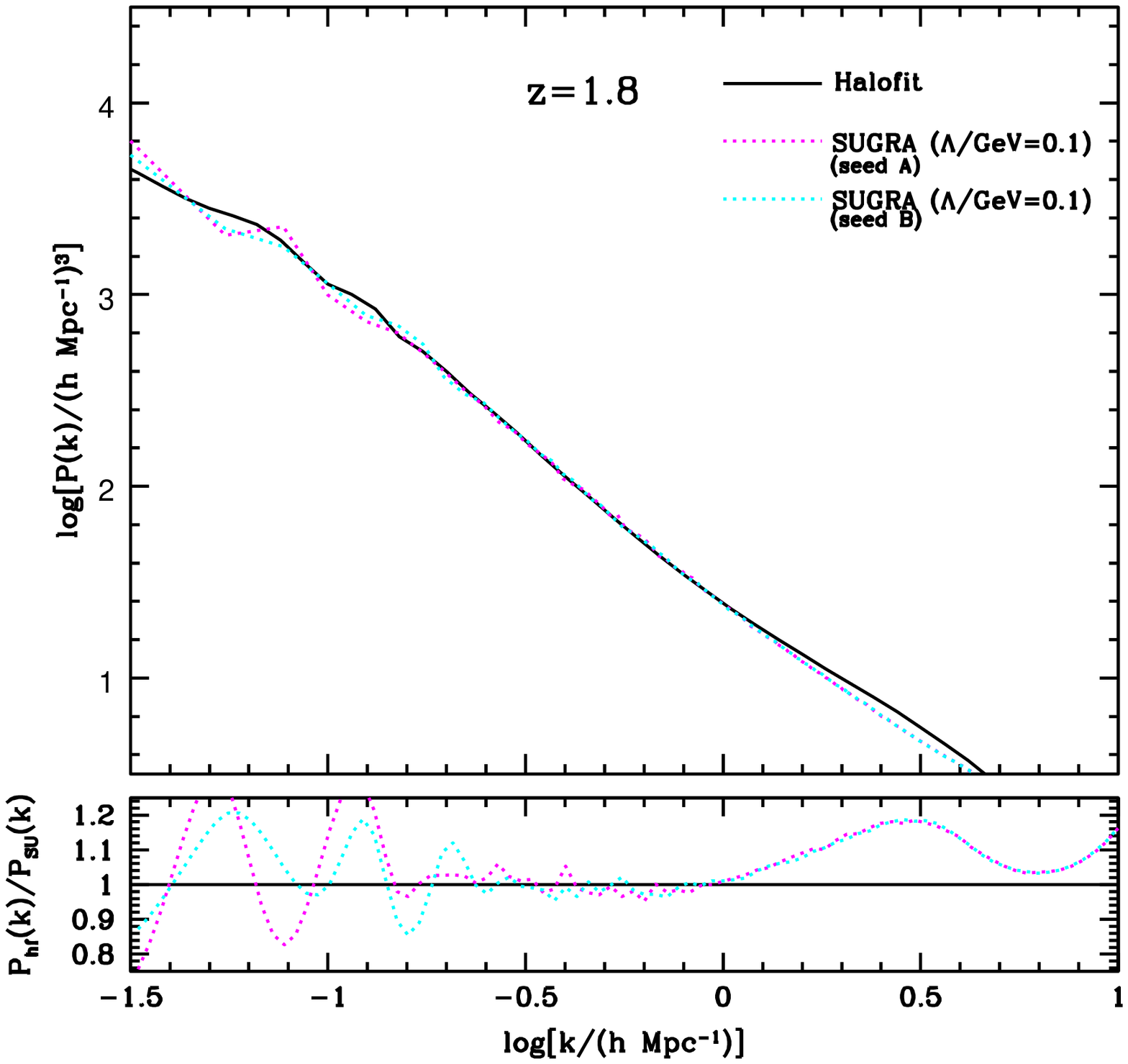}
\end{center}
\vskip -1.4truecm
\caption{As Fig.~\ref{s00s04}, for $z=1.2$ and $z=1.8~.$ Notice the
gradual deterioration of Halofit, which becomes unsuitable on the
whole non--linearity range.}
\label{s08s10}
\end{figure}
\begin{figure}
\begin{center}
\includegraphics*[width=11cm]{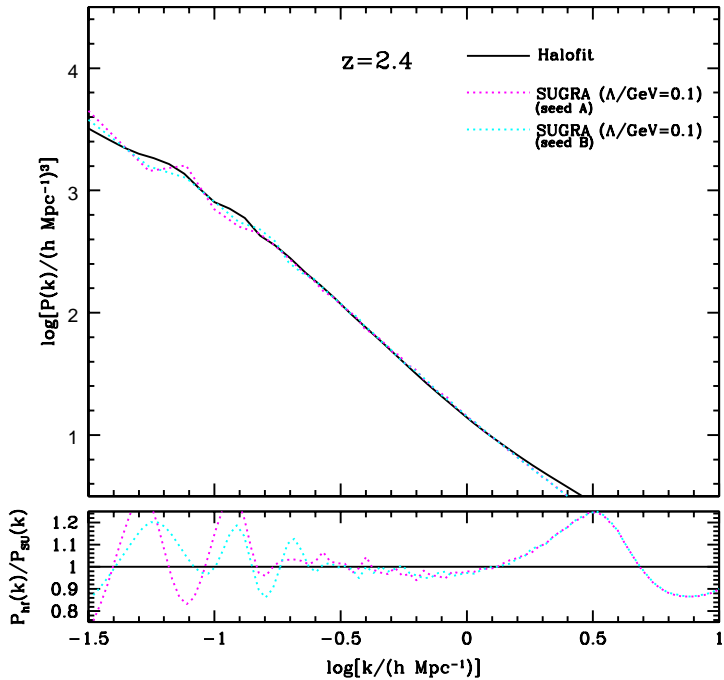}
\end{center}
\vskip -1.6truecm
\caption{As Fig.~\ref{s00s04}, for $z=2.4~.$
Discrepancies from Halofit exceed 20$\, \%~.$
}
\label{s24s10}
\end{figure}
\begin{figure}
\begin{center}
\vskip -7.truecm
\includegraphics*[width=11cm]{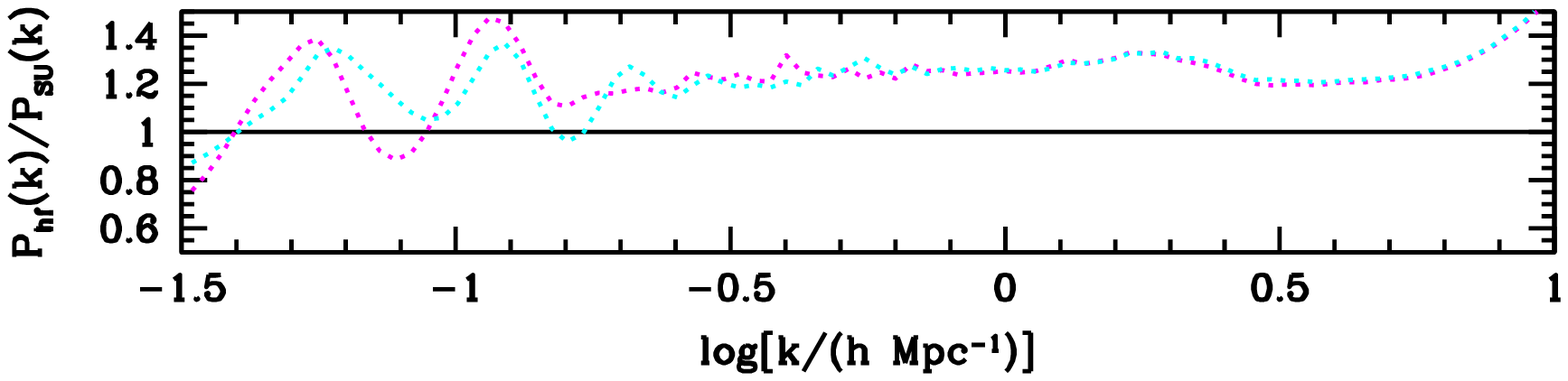}
\end{center}
\vskip -1.7truecm
\caption{Comparison of spectra with the fitting expression with a
correction for $w \neq -1$ \cite{mcdonald}, at $z=1.2$. Quite in
general, for our models, such expressions are no improvement in
respect to Halofit (notice that the ordinate range has doubled). This
is not unexpected, as they are used out of the allowed parameter
range}
\label{mcd}
\end{figure}

Power spectra have been computed from N--body simulations by using the
program PMpowerM of the ART package. The program works out the
spectrum through a FFT (Fast Fourier Transform) of the matter density
field, computed on a regular grid $N_G\times N_G\times N_G$ from the
particle distribution via a CIC (Cloud in Cell) algorithm.

Figure \ref{s00s04}, \ref{s08s10} and \ref{s24s10} present results on
the Power Spectrum extracted from N--body simulations of the SUGRA
model at $z=0$, 0.6, 1.2, 1.8 and 2.4~. Their main significance is to
allow a
comparison between seeds and with Halofit expressions.
The success of the {\it weak requirement} approach is so complete,
that differences between SUGRA and its $\cal W$ models cannot be
appreciated, even in the lower panel, where we report the ratio
between spectra from simulations and fitting formulae. The situation
is similar for the polynomial model (not plotted). Residual
differences between $\cal M$ and $\cal W$ models will be shown in a
following plot. Here, different colors refer to two different
realizations,

The first Figure \ref{s00s04} shows a rather good fit with Halofit
expressions at $z=0$. Here we have another problem, which must be
deepened by using simulation in a bigger bix and with a wider
dynamical range; in fact, the differences between realizations are $<
1\, \%$ only for $k > \sim 3/h{\rm Mpc}^{-1}$ (where hydrodynamics
becomes essential to describe the real world) and mostly keep within
2$\, \%$ for $k > \sim 0.6/h{\rm Mpc}^{-1}$. At still lower $k$'s, the
seed dependence becomes more and more significant. A tentative
explanation of the effect is that the relevant long wavelengths are
still not sampled well enough, in a box of the size considered in this
work, to yield a seed independent description at $z=0$, when spectral
contribution from long wavelengths had enough time to propagate down
to such high wavenumbers.

In fact, already at $z = 0.6$ seed discrepancies $> 1\, \%$ are
circumscribed to $k < \sim 0.6/h{\rm Mpc}^{-1}$. On the contrary, at
this redshift, the discrepancy from Halofit steadily exceeds 4--$5 \,
\%~.$

Higher redshift plots show a progressive deterioration of the Halofit
expressions, which become unsuitable in the whole non--linearity
range. No surprise about that, however, Halofit was built to meet
$\Lambda$CDM spectra and its performance in the cases considered here
is better than expected.

One could presume that using the expressions \cite{mcdonald}, aimed to
fit $w \neq -1$ cosmologies, could allow some improvement. As well as
Halofit, they are out of their range, but they include terms just
aimed to correcting for $w \neq -1~.$ Unfortunately, they yield no
improvement.
As an example, in Figure \ref{mcd}, we plot spectral ratios at $z =
1.2~$.


\begin{figure}
\begin{center}
\includegraphics*[width=11cm]{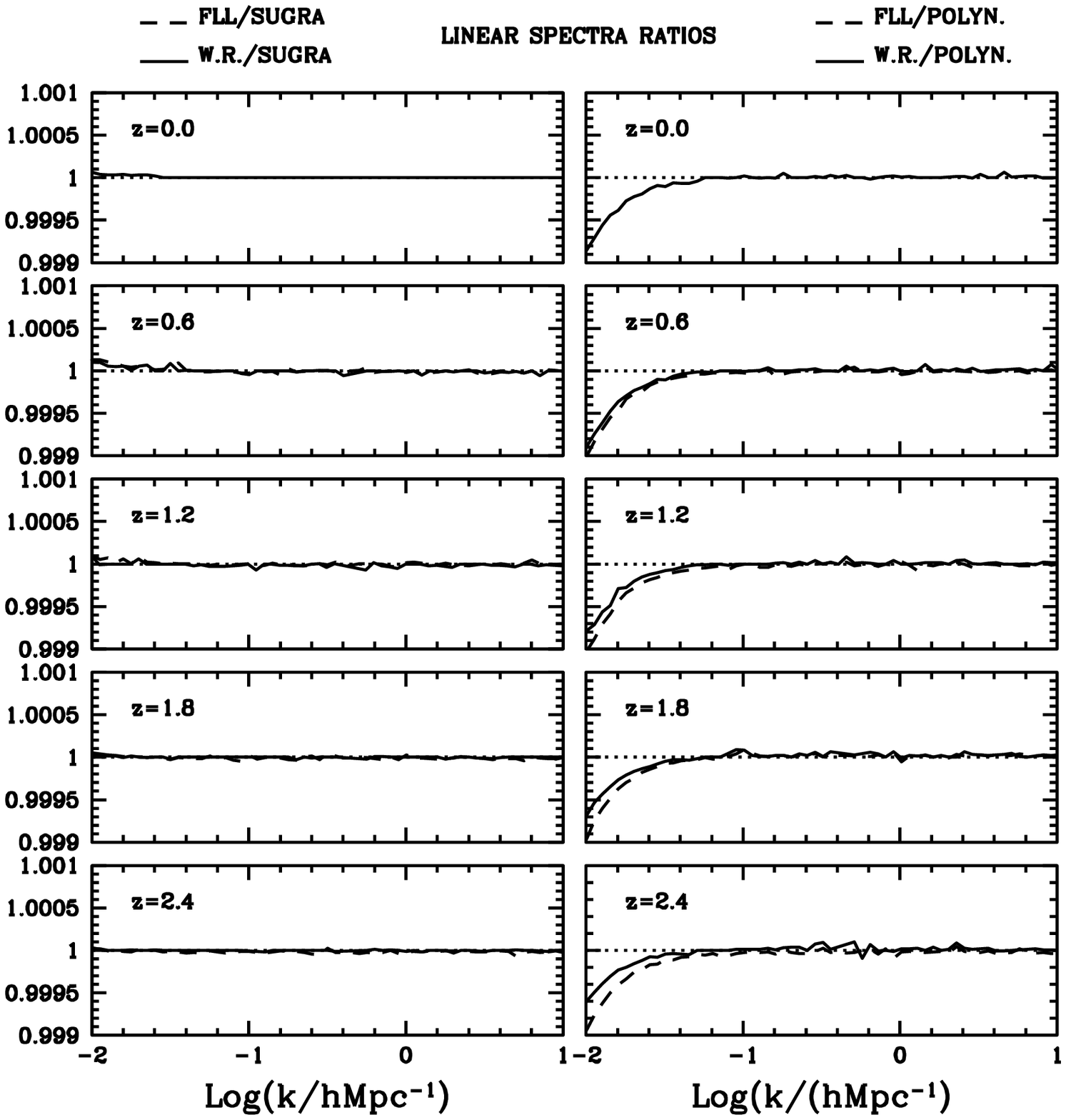}
\end{center}
\vskip -.1truecm
\caption{Linear spectra ratios.}
\label{s12}
\end{figure}

Let us now pass to the basic topic of this work and discuss
discrepancies between $\cal M$ and $\cal W$ models. To this aim, it is
useful to perform a preliminary comparison between the linear
spectra.  Figures \ref{s12} show spectral ratios for the polynomial
and SUGRA models at redshift values distant 0.6 up to z=2.4~. For the
polynomial model they can be confused with noise, apart of some
extremely mild signal at low $k$ ($\sim 1$--$2:100000$). Low $k$
discrepancies are greater (in the $k$ range considered they are $\sim
1:10000~$!) for SUGRA. Here we also appreciate a slight improvement
from the FLL to our approach: in the former case, all spectra concern
the same model, with the $w$ value deduced for $z=0$; in the latter
one, we use a $z$--dependent $w$. But, at this discrepancy level, the
main point to appreciate is that the linear spectra of $\cal W$$(z_k)$
are quite a good fit of the linear spectra of $\cal M$ at $z_k$.

\begin{figure}
\begin{center}
\includegraphics*[width=15cm]{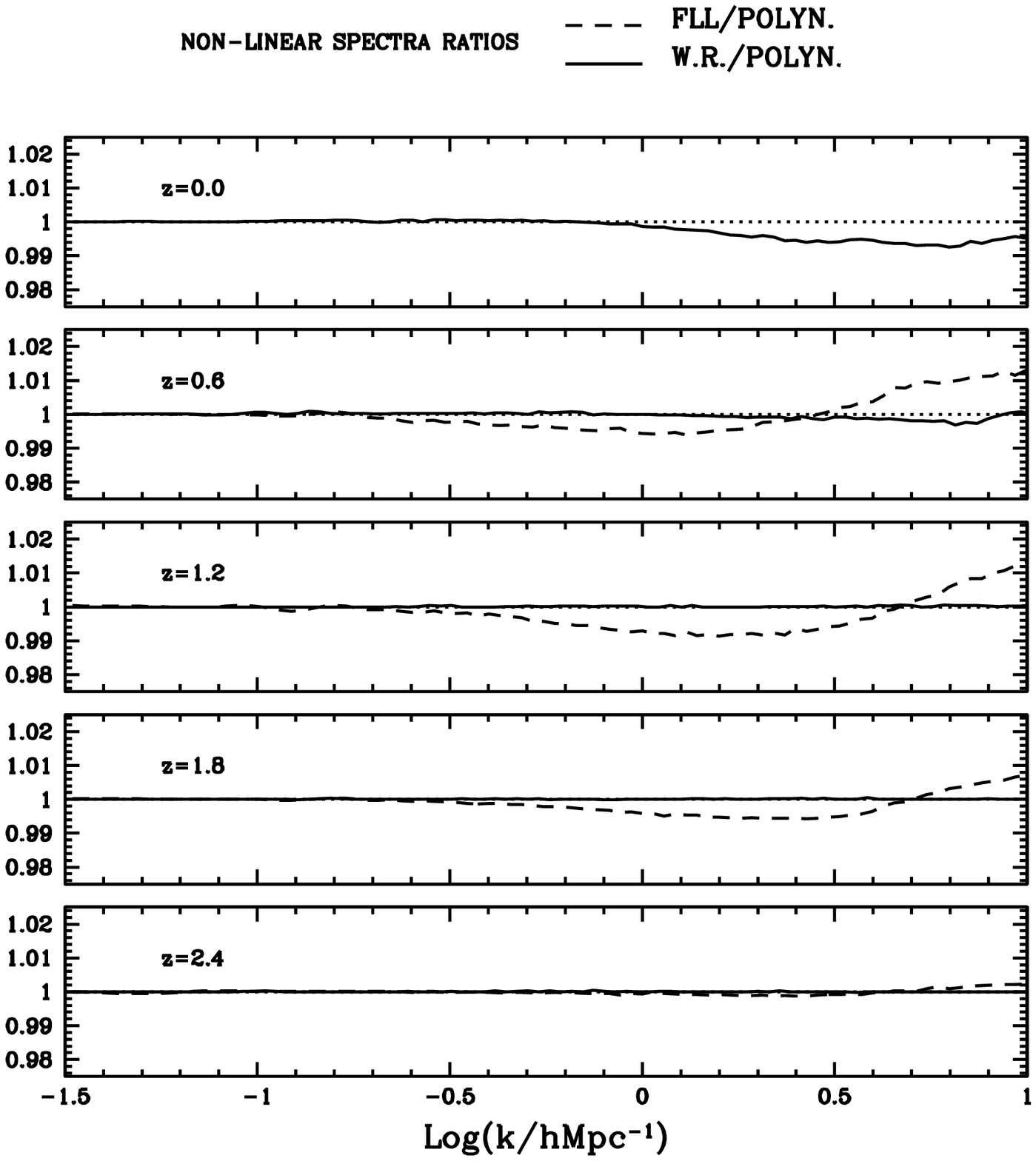}
\end{center}
\caption{Ratio between power spectra for the polynomial model and the
corresponding auxiliary models; dashed (solid) lines are obtained by
using FLL (our) technique.  Spectra for two model realizations are
plotted, but differences are just marginally appreciable for the
dashed lines.  Up to $k = 3\, h$Mpc$^{-1}$, the FLL approach already
provides results with discrepancies within 1$\, \%$.  They are
smallest at $z = 2.4$, a redshift close to crossover.  By using the
weak requirement (W.R.) technique described in this work, the maximum
discrepancies occur at $z=0$, where they are presumably caused by the
different non--linear history in $\cal M$ and $\cal W$ models,
propagating its effects even below $k = 3\, h$Mpc$^{-1}$. At $z=0.6$
there is a residual discrepancy peaking at a value $\sim 0.3\, \%$ on
a wavenumber $k \simeq 7\, h$Mpc$^{-1}$. Otherwise, discrepancies keep
within a fraction of permil, at any $k$.}
\label{PO}
\end{figure}

\begin{figure}
\begin{center}
\includegraphics*[width=15cm]{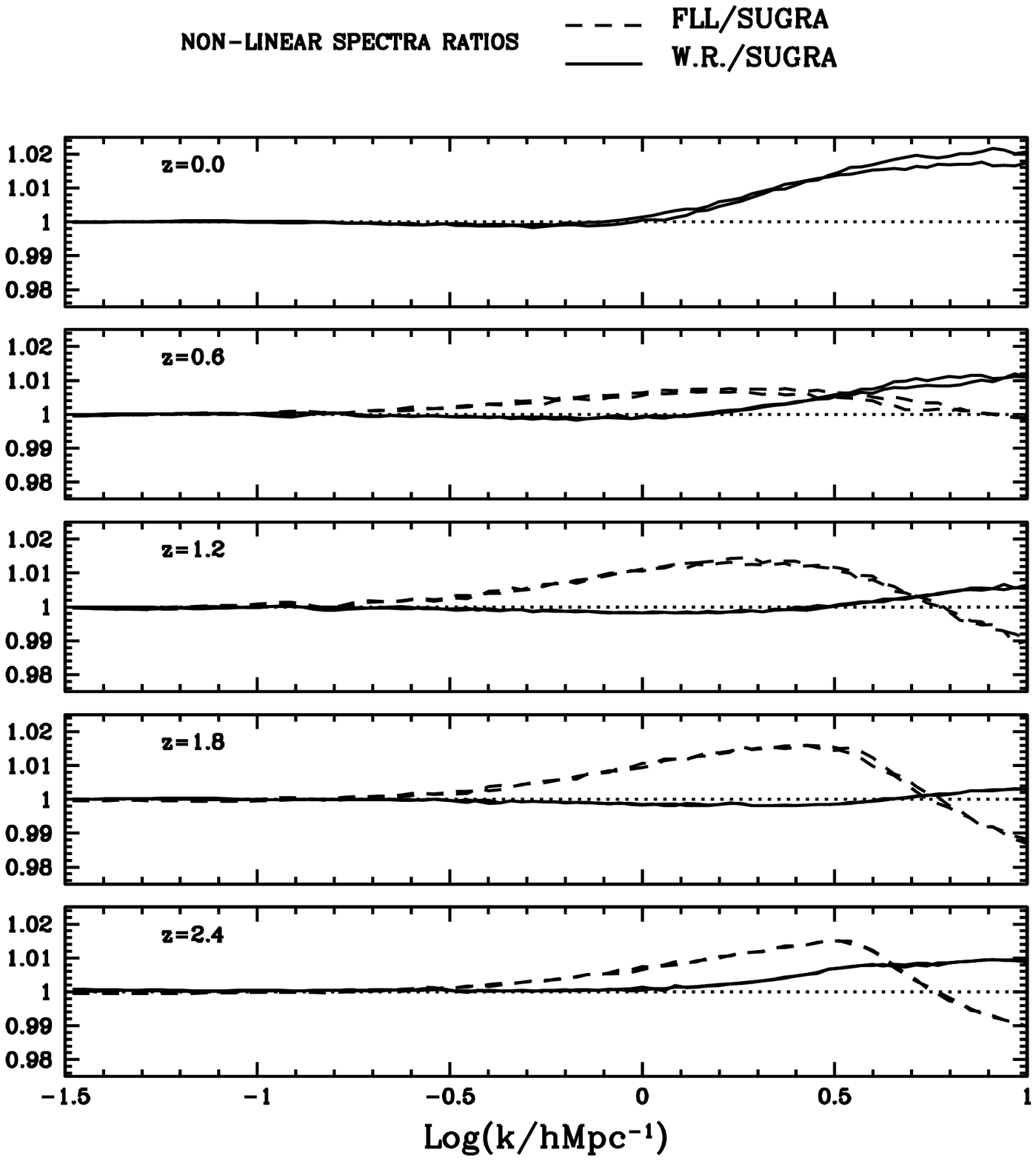}
\end{center}
\caption{Ratio between power spectra for the SUGRA model and the
corresponding auxiliary models; dashed and solid lines as in previous
Figure. Again spectra for two model realizations are plotted and some
slight differences are appreciable. At $z=0$, where FLL and
W.R. approaches coincide, differences between $\cal M$ and $\cal W$
models keep within $1 \%$ just up to $k = 1.8\, h$Mpc$^{-1}$; at $k =
3\, h$Mpc$^{-1}$, they are $\sim 1.3\, \%$. At higher redshift, the
FLL technique provides discrepancies within 1$\, \%$ just at $z=0.6$.
The W.R. technique here yields a great improvement, keeping
discrepancies within 0.6$\, \%$, 0.1$\, \%$ 0.2$\, \%$ 0.6$\, \%$ at
$z = 0.6$, 1.2, 1.8, 2.4, respectively.}
\label{SU}
\end{figure}

The main results of this work are then described in the Figures
\ref{PO} and \ref{SU}. Let us remind that the polynomial model is one of
those already considered by FLL, selected for the sake of comparison.
As could also be expected from the linear spectra, both techniques
perform better for this model, while the SUGRA model puts a more
severe challenge.

In Figure \ref{PO} we show the ratio between power spectra for the
polynomial model and the corresponding $\cal W$$(z_k)$ models.  The
spectra worked out by using the W.R. technique are given by solid
lines. We also plot those obtained according to FLL (dashed lines).
Although spectra for two model realizations are plotted, their
differences can hardly be appreciated.

Then, in Figure \ref{SU} we also show the ratio between power spectra
for the SUGRA model and the corresponding auxiliary models $\cal
W$$(z_k)$. Differences between model realizations are more easily
appreciable here, than in the Figure \ref{PO} , but still quite
negligible. What is immediately visible, instead, is the greater
difficulty that both W.R. and FLL techniques have to approach SUGRA
spectra.

Two different limitations are to be considered, when using such
techniques. The first one arises from hydrodynamics and related
effects. At $z=0$ hydrodynamics pollutes N--body spectra when
$k/h$Mpc$^{-1}$ overcomes $ \sim 3$: the number of halos with mass $>
(4\pi/3)~3^3 \times 2.78 \cdot 10^{11} \Delta_{vir} M_\odot h^{-1}$ is
enough to induce a spectral distortion $> 1\, \%$. At greater
redshift, the same number of halos can be found only on smaller
scales. Accordingly, N--body results are safe from hydro--pollution up
to an increasing value of $k$. It is sufficient to look at Figure
\ref{fmass} to appreciate the scaling of the limiting value, which (in
$h$Mpc$^{-1}$ units) will be $\sim 6$ at $z \simeq 1.2$ and $\sim 8$
at $z \simeq 2.4\, $.

The second limitation is the one we test here. Possible differences
between the spectra of the models $\cal M$ and $\cal W$$(z)$, at the
redshift $z$, can arise from the different $a(t)$ history in the two
models, after the non--linearity onset. Such differences shall be
greater on greater $k$ (smaller scales), which became non--linear
earlier. At a fixed $k$, differences in the non--linearity history
will become more and more significant at later times. 

These discrepancies are those we try to minimize and are shown in
Figures \ref{PO},~\ref{SU}~. In the $k$ range considered, they are
already practically absent at $z=0.3\, $, when using the W.R.
approach. This is shown by all plots for $z \neq 0$ in the above two
Figures. Here, the critical $k$ for hydro distortions is systematically
below the $k$ where the non--linear history begins to matter. Just at
$z \simeq 0$ the two critical $k$'s are close, their relative setting
being somehow model dependent.

Accordingly, while the spectra of the polynomial model, also treated
by FLL, are easily met by the relative $\cal W$ model at $z=0$, SUGRA
confirms to be a harder challenge. Figures \ref{PO} and \ref{SU}
however show that the W.R. technique is successful at any $z$, mostly
attaining a precision at the per mil level, and hardly exceeding the
1$\, \%$ discrepancy even in the most difficult cases.

\section{Conclusions}
\label{sec:disc}

By using the {\it ``weak requirement''} condition described in this
paper, we showed that even spectra of models with rapidly varying
$w(a)$ are easily obtainable from constant--$w$ spectra. 
We also verify that available {\it Halofit} expressions are not a
sufficient approximation, already at $z \simeq 0.6~$. Such expressions
were generalized to $w={\rm const.}$ models by \cite{mcdonald}, but
for a range of values of $\sigma_8$ which do not cover our case, tuned
on recent observational outputs.

In our opinion, this means that strong efforts are soon to be made to
provide generalized Halofit expressions, for constant--$w$ models,
effective for the whole parameter range that recent observations
suggest to inspect, and working up to reasonably high redshift.

Let us then suppose that such expressions are available and that new
observations provide direct information on density fluctuation spectra
at $N$ redshifts $z_k$ ranging from 0 to $z_N$. Such spectra should
then be fit to models, directly assuming that the reduced density
parameters
\begin{equation}
\omega_c (z_k) = \omega_{o,c} (1+z_k)^3 ~,~~~~~
\omega_b (z_k) = \omega_{o,b} (1+z_k)^3 ~,
\label{omegas}
\end{equation}
as it must however be, and seeking suitable $w_k$, $\sigma_8^{(z_k)}$,
$h_k$. Using then each $w_k,$ the $z=0$ values $\sigma_{o,8}^{(z_k)}$
and $h_{o,k}$ should be easily reconstructed. The $z_k$ dependence of
$w_k$, $\sigma_{o,8}^{(z_k)}$, $h_{o,k}$ should then allow to recover
a physical $w(z)$ behavior.

Let us outline, in particular, that the $w_k$ values directly measured
are not the physical $w(z)$. On the contrary, they can be quite far
from it, as is made clear in Figure \ref{wzwz}: there, the measured
$w_k$ correspond to the solid curves, while the physical state
parameter scale dependence is given by the dotted curves.

The simultaneous use of the information on $\sigma_{o,8}^{(z_k)}$
(fitting curves similar to those in Figure \ref{sigz2}, where it is
simply named $\sigma_8$) and $h_{o,k}$ (which can be easily determined
through linear programs), can allow a complete exploitation of
forthcoming data.

Preliminary evaluations on the possible efficiency of observational
techniques probing high redshift spectra, as the planned DUNE--EUCLID
experiment, should be reviewed and possibly improved on the basis of
the above conclusions.

\section*{Acknowledgments}

\noindent
Thanks are due to Anatoly Klypin and Gustavo Yepes for wide
discussions.  We are indebted with Joachim Stadel for granting us the
use of the {\sc pkdgrav} code. We are also indebted with Giuseppe La
Vacca for allowing us Figure 1, before its publication.  Most
numerical simulations were performed on the PIA cluster of the
Max-Planck-Institut f\"ur Astronomie at the Rechenzentrum in
Garching. Finally, it is a pleasure to thank an anonymous referee
whose suggestions allowed us to improve and complete the presentation
of our results. The support of ASI (Italian Space Agency) through the contract I/016/07/0 ``COFIS'' is acknowledged.

\appendix
\section{ {\sc pkdgrav} and Dynamical Dark Energy}

The central structure in {\sc pkdgrav} is a tree structure which forms
the hierarchical representation of the mass distribution.  Unlike the
more traditional oct-tree {\sc pkdgrav} uses a k--D tree, which is a
binary tree. The root-cell of this tree represents the entire
simulation volume. Other cells represent rectangular sub-volumes that
contain the mass, center-of-mass, and moments up to hexadecapole order
of their enclosed regions. {\sc pkdgrav} calculates the gravitational
accelerations using the well known tree-walking procedure of the
Barnes-Hut algorithm~\cite{barnes}.  Periodic boundary conditions are
implemented via the Ewald summation technique~\cite{hernqvist}.  {\sc
pkdgrav} uses the ordinary time $t$ (in suitable units) as independent
variable. The link between the expansion factor $a$ and the ordinary
time $t$, in the case ($\Lambda$)CDM models is based on the equation:
\begin{equation}
\dot a/a = H(t) = (H_0 a^{-2}) (\Omega_r + \Omega_m a +
\Omega_{\kappa} a^2 + \Omega_\Lambda a^4)^{1/2}
\label{Ha}
\end{equation}
where all the density parameters indicate redshift zero values.
Herefrom one obtains soon that
\begin{equation}
t = \int_0^a da/\{aH(a)\} = (2/3)\int_0^{Y(a)} dy/\{yH[a(y)]\}
\end{equation}
with
\begin{equation}
Y(a)=a^{3/2},~a(y)=y^{2/3}
\label{int}
\end{equation}
and the change of variable is clearly tailored on models based on CDM.
All this procedure is modified in our algorithm. The program now
creates {\it a priori} the $t(a)$ dependence, by integrating a
suitable set of differential equations. A large number ($\sim 10000$)
of $t(a)$ values are then kept in memory and interpolated to 
work out $t(a)$ at any $a$ value needed during the simulation run.

\vfill\eject
\section*{References}

\newcommand{\Nature}{{\it Nature\/} }
\newcommand{\ApJ}{{\it Astrophys. J.\/} }
\newcommand{\ApJS}{{\it Astrophys. J. Suppl.\/} }
\newcommand{\MNRAS}{{\it Mon. Not. R. Astron. Soc.\/} }
\newcommand{\PhRv}{{\it Phys. Rev.\/} }
\newcommand{\PhL}{{\it Phys. Lett.\/} }
\newcommand{\JCAP}{{\it J. Cosmol. Astropart. Phys.\/} }
\newcommand{\AeA}{{\it Astronom. Astrophys.\/} }
\newcommand{\etall}{{\it et al.\/} }
\newcommand{\arXiv}{{\it Preprint\/} }

\end{document}